\documentclass[aps,preprint,tightenlines,showpacs,nofootinbib]{revtex4}
\usepackage{epsfig}
\usepackage{multirow}
\usepackage{amsmath,stackrel}
\usepackage{amssymb}
\usepackage{graphicx}

\begin{document} 

\title{Heavy-baryon quark model picture from lattice QCD}
\author{J.~Vijande}
\affiliation{Departamento de F\'{\i}sica At\'{o}mica, Molecular y Nuclear, Universidad de Valencia (UV)
and IFIC (UV-CSIC), Valencia, Spain.}
\author{A.~Valcarce}
\affiliation{Departamento de F\'\i sica Fundamental, Universidad de Salamanca, E-37008
Salamanca, Spain}
\author{H.~Garcilazo}
\affiliation{Escuela Superior de F\'\i sica y Matem\'aticas,
Instituto Polit\'ecnico Nacional, Edificio 9,
07738 M\'exico D.F., Mexico}
\date{\emph{Version of }\today}

\begin{abstract}
The ground state and excited spectra of baryons 
containing three identical heavy quarks, $b$ or $c$,
have been recently calculated in nonperturbative lattice QCD. 
The energy of positive and negative parity excitations has been determined
with high precision. Lattice results constitute a unique opportunity to 
learn about the quark-confinement mechanism as well as elucidating our 
knowledge about the nature of the strong force.
We analyze the nonperturbative lattice QCD results by means of heavy-quark static 
potentials derived using SU(3) lattice QCD. We make use 
of different numerical techniques for the three-body problem.
\end{abstract}
\maketitle

\section{Introduction}
A precise calculation of the mass of the ground state
triply-bottom baryon, $\Omega_{bbb}$, and ten excited 
positive and negative parity states with angular momentum 
up to $J=7/2$ has been recently performed in nonperturbative 
lattice quantum chromodynamics (QCD) with $2+1$ flavors of 
dynamical light quarks in Refs.~\cite{Mei10,Mei12}.
The $b$ quark was implemented very accurately with 
improved lattice nonrelativistic QCD as suggested in Ref.~\cite{Tha91}.
There are also recent nonperturbative calculations of the 
ground and excited states with angular momentum 
up to $J=7/2$ of triply-charm baryons using anisotropic 
lattice QCD with a background of $2+1$ 
dynamical light quark fields~\cite{Pad13}.
Within the same framework of Refs.~\cite{Mei10,Mei12}, 
dynamical $2+1$ flavor lattice QCD and improved nonrelativistic lattice QCD 
for the $b$ quarks, the bottomonium spectrum was computed in Ref.~\cite{Men10}
obtaining an excellent agreement with the experiment.
A calculation of the charmonium spectrum with the same lattice action and the same 
lattice spacing used in Ref.~\cite{Pad13} can be found in Ref.~\cite{Liu12}, where one can get
an idea of the typical size of the systematic uncertainties.
Such studies grant the lattice QCD results for the bound state
problem of three-heavy quarks with the category of experimental
data and thus provide us with a unique opportunity to test phenomenological 
quark models for baryons in the energy regime where the description 
using potential models is expected to work best.

As pointed out by Bjorken time ago~\cite{Bjo85}, bound states of three-heavy quarks, $QQQ$,
may provide a new window for the understanding of baryon structure.
Baryons made of three-heavy quarks reveal a pure baryonic spectrum without 
light-quark complications and provide valuable insight into the quark-confinement 
mechanism as well as to elucidate our knowledge about the nature of the strong 
force. On the theoretical side one would expect the potential models would be able
to describe triply-heavy baryons to a similar level of precision as their success in
charmonium and bottomonium. In the same way the $Q\bar Q$ interactions are examined 
in heavy mesons, the study of triply-heavy baryons will probe the $QQ$ interactions in the 
heavy quark sector. Triply-heavy baryons have been studied by different methods, 
including quark models (see Refs.~\cite{Kle10,Cre13} for a comprehensive review 
of quark-model theoretical approaches), QCD sum rules~\cite{Zha09,Wan12,Ali13,Ali14} and potential
nonrelativistic QCD with static potentials from perturbation theory
at leading order~\cite{Jia06} and next-to-next-to-leading order~\cite{Bra05,Bra10}.
However, no experimental results are available so far for triply-heavy baryons (see
Ref.~\cite{Che11} for a recent calculation of production cross section at the LHC),
and thus, the predictions of their properties cannot yet be compared to the
real world. 

Our poor knowledge of the three-quark, $3Q$, potential stems on the difficulty to produce 
$QQQ$ states and, as said above, the consequent lack of experimental 
data. The precise calculation of the ground and excited states of
triply-bottom baryons~\cite{Mei10,Mei12}, together with the ground and excited states of 
triply-charm baryons~\cite{Pad13}, provide us with the ideal testbench
to improve our understanding of phenomenological quark models for baryons
in the heavy-quark sector. The quark-model dependent calculations could be 
tested by comparing them to nonperturbative first-principles calculations 
in lattice QCD of the $bbb$ and $ccc$ systems. An adequate starting point 
may be the static $3Q$ potential derived in SU(3) lattice QCD that will be
described in the next section. In Sect.~\ref{Comp} we will present our
results making emphasis in the comparison between nonperturbative lattice QCD
data and the results obtained with the static $3Q$ potential derived in SU(3) lattice QCD.
For this purpose we will make use of different numerical techniques for the three-body 
problem, generalized Gaussians variational approaches, hyperspherical
harmonics and Faddeev equations~\cite{Gar07,Val08,Vij09}, to test quark-quark
potential descriptions in the heavy quark limit.
Finally in Sect.~\ref{Res} we will briefly summarize the 
main conclusions of this study. 

\section{The static three-quark potential in SU(3) lattice QCD}

Since the early days of QCD the interaction among heavy quarks has been explored
as an important tool to learn about the behavior of 
QCD at low-energies. The $Q\bar Q$ static potential has been studied 
extensively by lattice gauge theories~\cite{Bal01}, being nowadays
a very well-known quantity. The typical shape of the
color-singlet $Q \bar Q$ static potential, which is characterized by
a short-range Coulomb behavior and a long-range linear
rise, well represents the double nature of QCD as an
asymptotically free and infrared confined theory. 
The excitation spectrum of the gluon field around 
a static quark-antiquark pair has also been explored 
by lattice calculations~\cite{Jug03}.
On the large length scale the spectrum agrees with 
that expected for stringlike excitations while
in the short range it shows a Coulomb-like behavior 
as it was first noted within the context of the static 
bag picture of gluon excitations~\cite{Has80}.

In QCD the three-quark potential is of prime importance
reflecting the SU(3) gauge symmetry and being directly responsible for
the structure and properties of baryons, in the same way the $Q \bar Q$
potential is responsible for the meson properties. Furthermore,
the $3Q$ potential is a key quantity to clarify the quark confinement
mechanism in baryons. However, up to now
lattice QCD studies have paid relatively less attention to the potential 
that describes the interaction of three-heavy quarks,
as a consequence of their still missing experimental evidence.
Thus, the $3Q$ potential is much less known than the heavy $Q\bar Q$
potential for which many lattice studies exist~\cite{Bal01}. Even for the ground-state,
its accurate measurement using lattice QCD is relatively difficult
and has been performed recently~\cite{Tak01,Tak02,Ale02,Ale03}.
The gluonic excitation (excitations of the gluon field that are not thus of quark origin)
of the three-quark system has also been studied 
in SU(3) lattice QCD~\cite{Tak03,Tak04}, concluding that the lowest excitation
would be almost 1 GeV above the ground state.
This is rather large in comparison with the low-lying excitation
energy of the quark origin what makes the gluonic excitation
mode invisible in the low-lying excitations of baryons.
Thus, quark-degrees of freedom play the dominant role in low-lying baryons
with excitation energy below 1 GeV.
This large gluonic excitation energy is conjectured to give a 
physical reason of the success of the quark model for low-lying 
baryons even without explicit gluonic modes~\cite{Tak04}. 

Most of the existing lattice studies of the three-quark
static potential have mainly explored the region of large interquark
distances~\cite{Tak01,Tak02,Ale02,Ale03,Tak03,Tak04,Som86,Sug01,Bor04,Bon04,Cor04,Hub08,Iri11}.
As for the $Q\bar Q$ case, the characteristic signature of the 
long-range non-Abelian dynamics is believed to be a linear rising of 
the static interaction. Moreover, the general expectation for the baryonic
case is that, at least classically, the strings meet at the 
so-called Fermat (or Torricelli) point, which has minimum
distance from the three sources ($Y-$shape configuration). If
this is the case, one should see a genuine three-body
interaction among the static quarks. 
The confining short-range $3Q$ potential could be also 
described as the sum of two-body potentials 
($\Delta-$shape configuration)~\cite{Tak04,Ale03,Bor04,Cor04}. 
Many of the lattice calculations have focused 
on distinguishing the $Y-$ from the $\Delta-$configuration, despite 
the difference between a $\Delta$ and a $Y-$shape potential 
being rather small and difficult to detect.

In recent years, according to the remarkable progress of the computational power,
the lattice QCD Monte Carlo calculations have become a reliable and useful method for the
analysis of nonperturbative QCD. In particular, the $Q\bar Q$ potential, responsible
for the meson properties, has been extensively studied using lattice QCD.
The data of the $Q\bar Q$ ground state potential are well reproduced by a sum of a Coulomb term,
due to the perturbative one-gluon exchange (OGE), and a linear confinement~\cite{Tak02,Tak04},
\begin{equation}
V_{Q\bar Q}(r) = -\frac{A_{Q\bar Q}}{r} \, + \, \sigma_{Q\bar Q} \, r \, + \, C_{Q \bar Q} \,  .
\label{PotQQ}
\end{equation}
We give in Table~\ref{tab1} the parameters of the $Q\bar Q$ potential taken from Ref.~\cite{Tak02}.
The value of the $Q\bar Q$ confinement strength was determined to reproduce the value obtained
from the linear Regge trajectories of the pseudoscalar $\pi$ and $K$ mesons, $\sqrt{\sigma}=$
(429$\pm$2) MeV~\cite{Bal01}.
\begin{table}[t]
\caption{Standard string tension, $\sigma$, Coulomb coefficient, $A$, and constant term, $C$, of the static heavy 
quark potential obtained from SU(3) lattice QCD taken from Ref.~\cite{Tak02} for a lattice spacing $a=$ 0.19 fm.}
\label{Tab1}
\begin{center}
\begin{tabular}{cp{0.5cm}cp{0.5cm}cp{0.5cm}cp{0.5cm}cp{0.5cm}c}
\hline
             & & $\sigma$ ($a^{-2}$) & & $ \sigma$ (GeV$^2$)  & &   $A$       & & $C$ ($a^{-1}$) & &  $C$ (GeV)  \\ \hline
$Q\bar Q$    & & 0.1629(47)          & &  0.1757              & & 0.2793(116) & & 0.6203(161)    & & 0.6442       \\
$3Q(Y)$      & & 0.1524(28)          & &  0.1644              & & 0.1331(66)  & & 0.9182(213)    & & 0.9536       \\
$3Q(\Delta)$ & & 0.0858(16)          & &  0.0925              & & 0.1410(64)  & & 0.9334(210)    & & 0.9694       \\
\hline
\end{tabular}
\end{center}
\label{tab1}
\end{table}

However, there is almost no reliable formula to describe the $3Q$ potential
directly based on QCD, in spite of its importance for the study of baryon
properties. There have been recent advances on the determination of the ground state 
$3Q$ potential that is expected to take the form~\cite{Tak02},
\begin{equation}
V_{3Q}(r) = - A_{3Q} \sum_{i < j}\frac{1}{|\vec{r}_i - \vec{r}_j|} \, + \, \sigma_{3Q}^Y \, L_{\rm{min}} \, + \, C_{3Q} \,  ,
\label{Pot3QY}
\end{equation}
where $L_{\rm{min}}$ is the minimal value of the total length of color flux tubes
linking the three quarks, and
\begin{equation}
A_{3Q} \simeq \frac{1}{2}A_{Q\bar Q}\, , \,\,\,\,\,\,\,
\,\,\,\,\,\,\, \sigma_{3Q}^Y \simeq \sigma_{Q\bar Q} \, .
\end{equation} 
The short-distance behavior of $V_{3Q}(r)$ is expected to be described by the 
two-body Coulomb potential as the one-gluon exchange result in perturbative QCD.
Fit analysis in terms of a Yukawa potential considering a possible gluon mass $m_B$
have concluded $m_B=0$, what reduces the Yukawa to the Coulomb potential~\cite{Tak02}.
The one-gluon exchange result indicates also the simple relation on the Coulomb
coefficients in the $Q\bar Q$ and the $QQQ$ potentials as $A_{3Q} \simeq \frac{1}{2}A_{Q\bar Q}$.
For the long-distance behavior, the confining baryonic static potential
rises like the $Y-$ansatz~\cite{Tak04,Bor04}. 

Calculations both in full and quenched QCD demonstrates that the confining baryonic static
potential approaches the $\Delta-$ansatz at short distances~\cite{Ale03,Tak04,Bor04}.
The $\Delta-$ansatz behavior at short distances is of great importance for
phenomenological models since the calculation of orbital excited states
with the $\Delta-$ansatz are much more simpler.
For an equilateral $QQQ$ arrangement, as expected for a system
of three-heavy identical quarks, departure from the
$\Delta-$ansatz is not significant until $d_{QQ}\sim 0.7$ fm~\cite{Ale03,Bor04}, so
that the $\Delta-$ansatz may be the more relevant one
for quarks confined inside a baryon and then, for the case of $QQQ$ bound
states whose root-mean square radii are much smaller than 
such distance (see Table~\ref{tab2}).
Thus, the static three-quark potential could be described by
a simple sum of the effective two-body $QQ$ potentials
with a reduced string tension~\cite{Tak02,Ale02,Ale03},
\begin{equation}
V_{3Q}(r) = - A_{3Q} \sum_{i < j}\frac{1}{|\vec{r}_i - \vec{r}_j|} \, + \, \sigma_{3Q}^\Delta 
\sum_{i < j}|\vec{r}_i - \vec{r}_j| \, + \, C_{3Q} \,  ,
\label{Pot3QD}
\end{equation}
where
\begin{equation}
A_{3Q}\simeq \frac{1}{2}A_{Q\bar Q}\, , \,\,\,\,\,\,\, \sigma_{3Q}^\Delta \simeq 0.53 \sigma_{Q\bar Q}\, .
\end{equation} 
The reduction factor in the string tension can be naturally understood as a geometrical
factor rather than the color factor, due to the ratio between $L_{\rm {min}}$ and the 
perimeter length of the $3Q$ triangle, suggesting 
$\sigma_{3Q}^\Delta= \left(0.50 \sim 0.58 \right) \sigma_{Q\bar Q}$~\cite{Tak02}.
For the particular case of quarks in an equilateral triangle 
$\sigma_{3Q}^\Delta= \frac{1}{\sqrt{3}} \sigma_{Q\bar Q} = 0.58 \sigma_{Q\bar Q}$~\cite{Bor04}.
When the $\Delta-$ansatz is adopted for the
two-body linear potential, still the same relation holds for the strength of the Coulomb potential
$A_{3Q} \simeq \frac{1}{2}A_{Q\bar Q}$. This $\Delta-$ansatz has been widely adopted in the nonrelativistic
quark model because of its simplicity~\cite{Kle10,Cre13,Gar07,Val08,Isg78,Oka81,Sil96}

\section{Results and discussion}
\label{Comp}

In order to check if the static three-quark potential of Eq.~(\ref{Pot3QD}) (or Eq.~(\ref{Pot3QY}))
with the parameters in Table~\ref{tab1} determined from lattice QCD~\cite{Tak02} can reproduce the
$bbb$ and $ccc$ baryon spectra measured also in lattice QCD~\cite{Mei10,Mei12,Pad13} we will make
use of three different numerical methods: generalized Gaussians variational approach~\cite{Vij09}, 
hyperspherical harmonics~\cite{Val08} and Faddeev equations~\cite{Gar07}. The three-methods have
been used and the difference in results is negligible.
In all cases we solve the nonrelativistic Schr\"odinger equation
\begin{equation}
\left\{H_0 + V_{3Q}(r)\right\}\Psi(\vec r) = E \Psi({\vec r}) \, ,\nonumber
\end{equation}
where $H_0$ is the free part of three-heavy quarks without center of mass motion
\begin{equation}
H_0=\sum_{i=1}^3 \left( M_{Q} + \frac{\vec{p}_i^{\,2}}{2M_{Q}} \right) - T_{CM}\, \nonumber
\end{equation}
and $M_{Q}$ is the mass of the heavy quark. The mass of the heavy baryon will be finally 
given by $M_{3Q}=3M_Q + E$.
\begin{table}[t]
\caption{Mass, in GeV, and root-mean square radius, in fm, of the ground state, $E_1(\frac{3}{2}^+)$, and the first 
positive parity excited state, $E_2(\frac{3}{2}^+)$, of $QQQ$ baryons for different values of the mass of the heavy quark 
$M_Q$, in GeV, and for the $Y-$shape and $\Delta -$shape confining potentials of Eqs.~(\ref{Pot3QY_conf})
and~(\ref{Pot3QD_conf}), respectively.}
\begin{center}
\begin{tabular}{|c|p{0.5cm}cp{0.5cm}cp{0.5cm}|p{0.5cm}cp{0.5cm}cp{0.5cm}cp{0.5cm}cp{0.5cm}|}
\hline
   $M_Q$  & & \multicolumn{3}{c}{$Y-$shape} && & \multicolumn{8}{c|}{$\Delta-$shape} \\
          & &  $E_1(\frac{3}{2}^+)$ & & $E_2(\frac{3}{2}^+)$ && & $E_1(\frac{3}{2}^+)$  & & $\sqrt{<r_1^2>}$ & & $E_2(\frac{3}{2}^+)$&  &$\sqrt{<r_2^2>}$  & \\ \hline
    4.0   & &  12.814     & &  13.121     && & 12.795     & &  0.236   & &   13.094 & &    0.334     &  \\
    4.4   & &  13.989     & &  14.286     && & 13.970     & &  0.228   & &   14.260 & &    0.324     &  \\
    4.8   & &  15.166     & &  15.455     && & 15.148     & &  0.222   & &   15.430 & &    0.315     &  \\
    5.2   & &  16.346     & &  16.627     && & 16.328     & &  0.216   & &   16.603 & &    0.306     &  \\
    5.6   & &  17.528     & &  17.801     && & 17.509     & &  0.211   & &   17.778 & &    0.299     &  \\
\hline
\end{tabular}
\end{center}
\label{tab2}
\end{table}

We have checked the validity of the $\Delta-$ansatz for the spectroscopy of triply-heavy 
baryons calculating their masses and root mean square radii by means of a simple confining 
interaction given either by a $Y-$shape or a $\Delta-$shape interaction. 
The $Y-$shape potential would be given by~\cite{Tak02},
\begin{equation}
V_{3Q}(r) = \sigma_{3Q}^Y \, L_{\rm{min}} \,  ,
\label{Pot3QY_conf}
\end{equation} 
where $L_{\rm{min}}$,
\begin{equation}
L_{\rm{min}}=\left[\frac{1}{2}\left( a^2+b^2+c^2 \right) +
\frac{\sqrt{3}}{2}\sqrt{\left(a+b+c\right)\left(-a+b+c\right)
\left(a-b+c\right)\left(a+b-c\right)}\right]^{1/2} \, ,
\end{equation}
being $a$, $b$ and $c$ the three sides of the $3Q$ triangle, when the angles of the
$3Q$ triangle do not exceed $2\pi/3$. When an angle of the $3Q$ triangle exceeds
$2\pi/3$, one gets,
\begin{equation}
L_{\rm{min}}= a + b + c - {\rm{max}}\left(a,b,c\right)\, .
\end{equation}
We have taken for the string tension a standard value 
of $\sigma_{3Q}^Y=$ 0.1648 GeV$^2$~\cite{Tak01,Tak02,Ale02,Ale03,Tak03,Tak04}. The $\Delta-$shape 
potential would have the form,
\begin{equation}
V_{3Q}(r) =  \sigma_{3Q}^\Delta \sum_{i < j}|\vec{r}_i - \vec{r}_j| \,  ,
\label{Pot3QD_conf}
\end{equation}
with $\sigma_{3Q}^\Delta= 0.53 \sigma_{3Q}^Y=$ 0.0874 GeV$^2$.
The results are shown in Table~\ref{tab2} for different values
of the heavy quark mass. As can be seen the two geometrical configurations give the same 
result with a small difference of 0.15\%. If the geometrical factor in front of
the $\Delta-$shape interaction is slightly modified, 0.55 instead of 0.53,
the $Y-$shape and the $\Delta-$shape results would be exactly the same.
This result was already noted in Ref.~\cite{Vij12}.
Besides, one observes how the root-mean square radii are much smaller 
than the quark-quark distance for which the $Y-$ and $\Delta-$shapes
start to slightly differ, $\sim$ 0.7 fm~\cite{Ale03}.
\begin{figure}[t]
\vspace*{-7cm}
\hspace*{-1cm}\mbox{\epsfxsize=180mm\epsffile{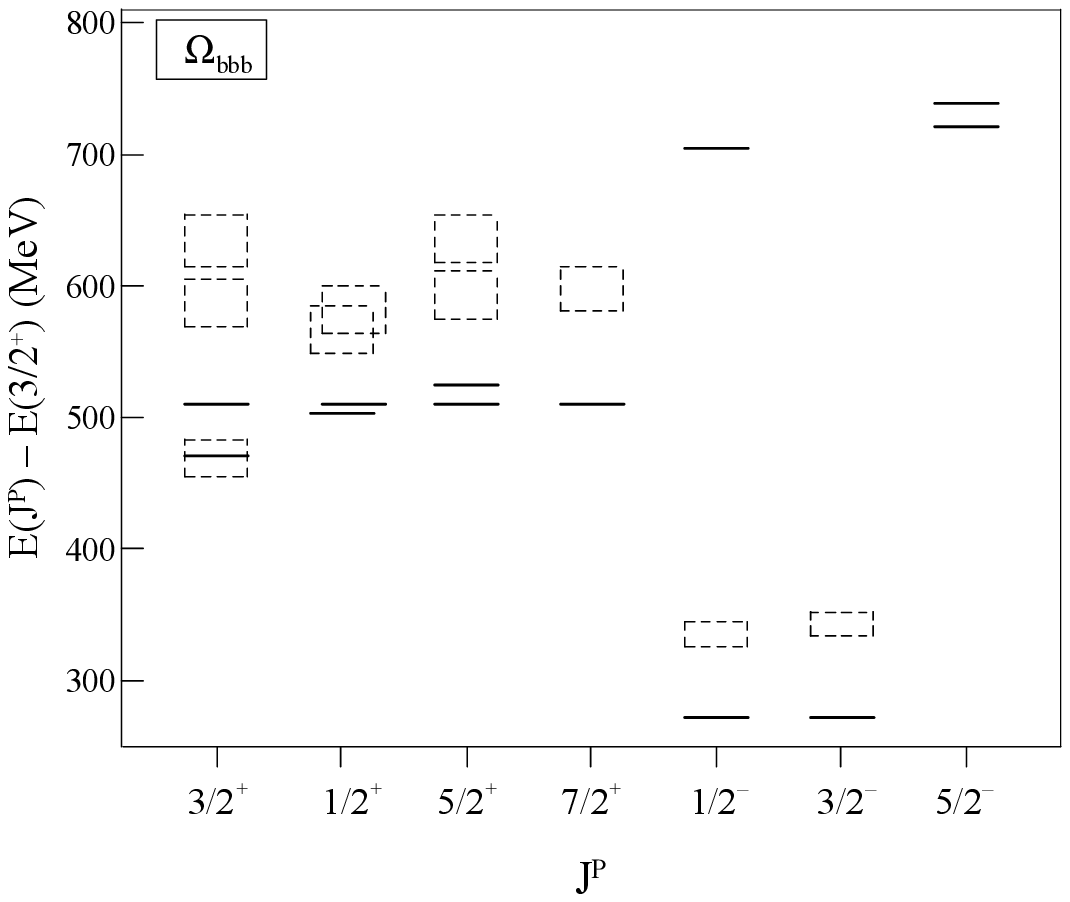}}\\
\vspace*{-17.cm}
\hspace*{-1cm}\mbox{\epsfxsize=180mm\epsffile{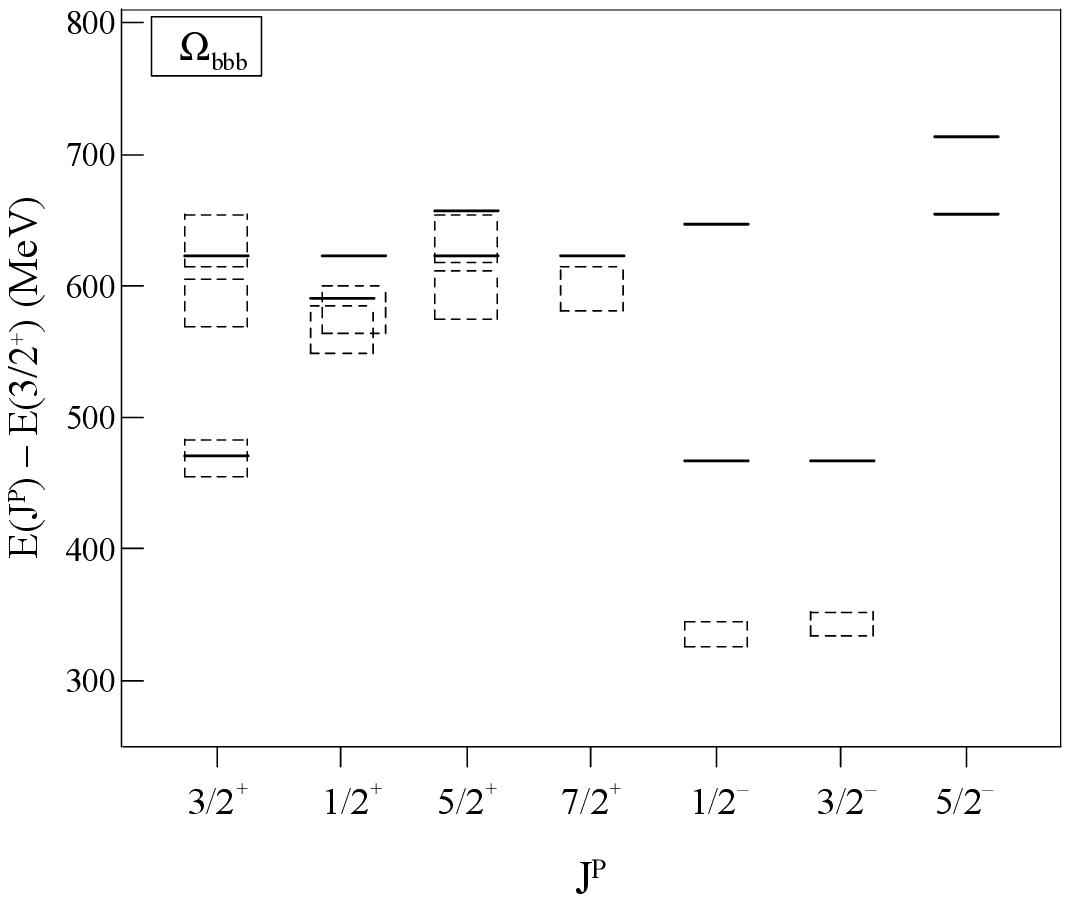}}
\vspace*{-12.cm}
\caption{$bbb$ excited state spectra, solid lines, for a single $\Delta-$shape confining 
potential (upper panel) or a single Coulomb interaction (lower panel). 
In both cases we have fixed the strength of the potential to reproduce the 
$E_2(3/2^+)-E_1(3/2^+)$ mass difference correctly, obtaining $\sigma_{3Q}^{\Delta}=$ 0.2076 GeV$^2$ 
for the case of the single $\Delta-$shape confining 
interaction and $A_{3Q}=$ 0.4231 for the case of the single Coulomb potential.
The boxes stand for the nonperturbative lattice QCD results of Ref.~\cite{Mei12}.}
\label{fig1}
\end{figure}

Looking for the general pattern of the results of nonperturbative lattice QCD for the
$bbb$ system we have compared against the result of a single $\Delta-$shape confining 
potential or a single Coulomb interaction. The results are shown in Fig.~\ref{fig1}.
In both cases we have fixed the strength
of the potential to reproduce the $E_2(3/2^+)-E_1(3/2^+)$ mass difference correctly, obtaining
$\sigma_{3Q}^{\Delta}=$ 0.2076 GeV$^2$ for the case of the single $\Delta-$shape confining 
interaction and $A_{3Q}=$ 0.4231 for the single Coulomb potential case.
We clearly note the large strength of the two interactions as compared to the
predictions of SU(3) lattice QCD (see Table~\ref{tab1}), where both terms would simultaneously
contribute. The $\Delta-$shape interaction (upper panel) produces a 
splitting between the positive (i.e., the first excited state of $J^P=3/2^+$) and 
negative (i.e., the first excited state of $J^P=3/2^-$) parity excited
states too large and a rather small excitation energy for the positive parity 
excitations. On the contrary, the Coulomb interaction (lower panel), being an almost hypercentral potential,
drives to the expected degeneracy between the positive and negative parity excited states,
not observed in the nonperturbative lattice QCD results. Thus, as the strength of the potential has been fixed
to reproduce the $E_2(3/2^+)-E_1(3/2^+)$ mass difference correctly,
the negative parity excited states are obtained close to the positive parity excited states in clear disagreement
with the nonperturbative lattice QCD results. Besides, the positive parity states (except the first excited state
of $J^P=3/2^+$ that has been fitted) are predicted slightly above the nonperturbative lattice QCD results.

\begin{figure}[t]
\vspace*{-7cm}
\hspace*{-1cm}\mbox{\epsfxsize=180mm\epsffile{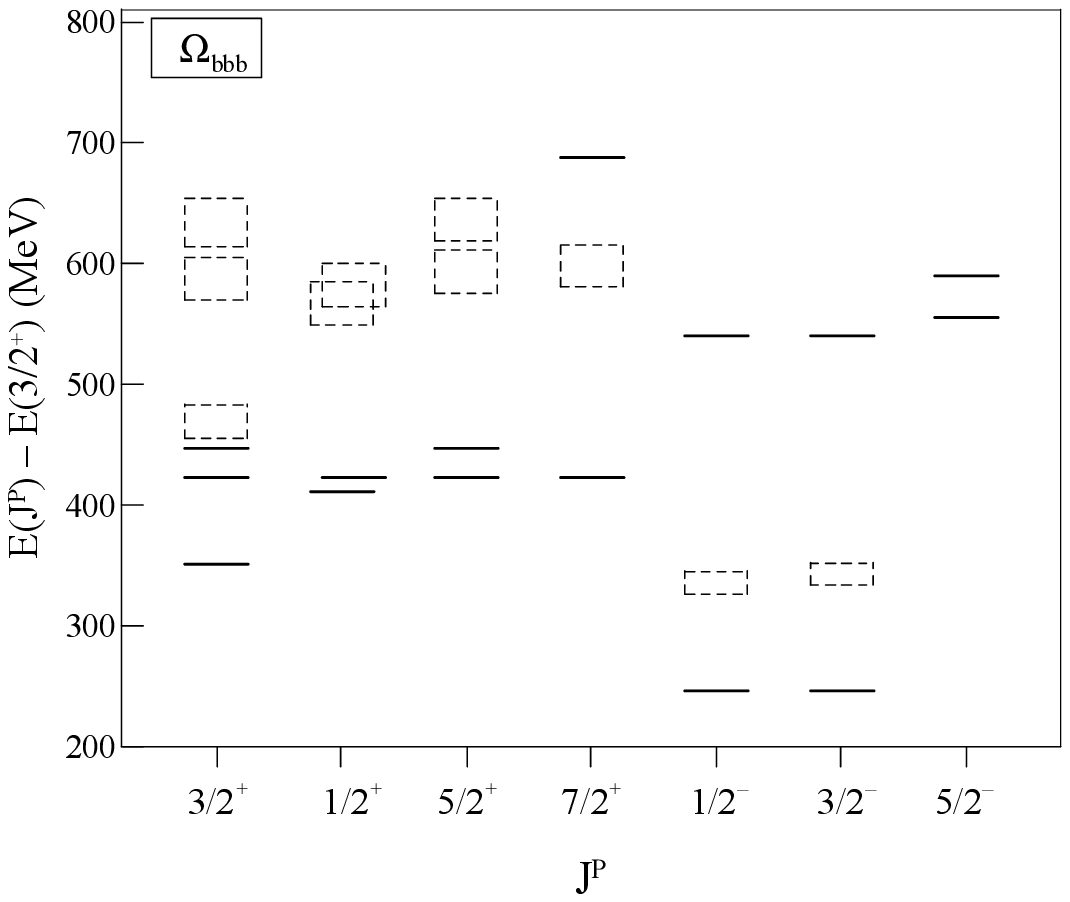}}\\
\vspace*{-17.cm}
\hspace*{-1cm}\mbox{\epsfxsize=180mm\epsffile{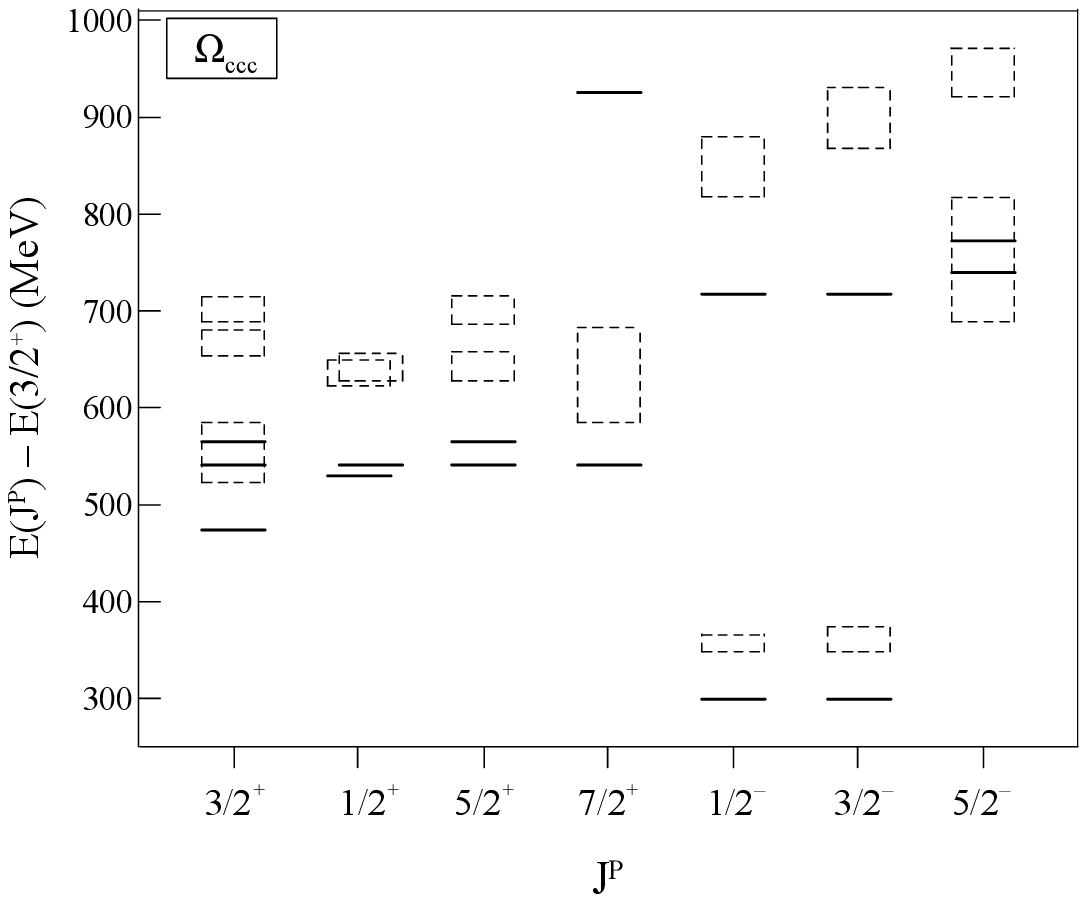}}
\vspace*{-12.cm}
\caption{$bbb$ (upper panel) and $ccc$ (lower panel) excited state spectra, solid lines, 
for the $3Q$ static SU(3) lattice QCD potential of Eq.~(\ref{Pot3QD}) with the parameters
of Table~\ref{tab1}. The boxes stand for the nonperturbative lattice QCD results 
of Ref.~\cite{Mei12} for the $bbb$ system and Refs.~\cite{Pad13,Mat14} for the $ccc$ system.}
\label{fig2}
\end{figure}

Thus, $bbb$ nonperturbative lattice QCD calculations point to a static $3Q$ potential
given by a mixture of a $\Delta-$shape confinement and a Coulomb interaction.
We have therefore make use of the standard $3Q$ static potential derived from SU(3) lattice QCD
and shown in Eq.~(\ref{Pot3QD}) with the parameters reported in Ref.~\cite{Tak02} and quoted
in Table~\ref{tab1}, $A_{3Q}=$ 0.1410 and $\sigma_{3Q}^\Delta=$ 0.0925 GeV$^2$ to calculate the
$bbb$ and $ccc$ spectra. The results are shown in Fig.~\ref{fig2}. We have fixed the 
mass of the heavy quark to reproduce the $J^P=3/2^+$ ground state of the $bbb$ and $ccc$ systems,
14.372 GeV~\cite{Mei10,Mei12} and 4.758 GeV~\cite{Pad13}, respectively, obtaining $m_b=$ 4.655 GeV
and $m_c=$ 1.269 GeV. As we can see in Fig.~\ref{fig2},
there is a large difference in the excited states
with the nonperturbative lattice QCD results, predicting a small splitting between positive and negative parity excited states
and also a small excitation energy for the positive parity states. These results clearly point to a lack of strength 
either in the confining or in the Coulomb potential.

One may think about several possible reasons for the disagreement: 
\begin{itemize}
\item The parameters of Eq.~(\ref{Pot3QD})
obtained in quenched QCD in Ref.~\cite{Tak02} might
be different from those in 2+1 flavor QCD employed in Refs.~\cite{Mei10,Mei12,Pad13}.
To check this possibility, one has to calculate parameters of Eq.~(\ref{Pot3QD})
in 2+1 flavor QCD but such calculations do not exist so far.
\item Results from both Ref.~\cite{Tak02} and Refs.~\cite{Mei10,Mei12,Pad13} contain systematic errors
such as lattice artifact and chiral extrapolation as well as statistical
errors. This may cause the disagreement.
\item The fitting form Eq.~(\ref{Pot3QD}), or Eq.~(\ref{Pot3QY}), might not be appropriate to describe
the static three-quark potential in lattice QCD.
\item The static quark description might be inaccurate for $bbb$  and, in
particular, for $ccc$.  Higher order terms might be necessary.
\item The quark model description with "3-quark potential" might be
inappropriate for $bbb$ and $ccc$ systems.
\end{itemize}

\section{Summary}
\label{Res}

To summarize, the spectra of baryons containing three identical heavy quarks, $b$ or $c$,
have been recently calculated in nonperturbative lattice QCD. 
The energy of the lowest positive and negative parity excited states has been determined
with high precision. These achievements constitute a unique opportunity to 
test phenomenological potential models in the regime where they are expected 
to work best. We have analyzed these results by means of static three-quark 
potentials derived using SU(3) lattice QCD using different numerical techniques 
for the three-body problem. 
Our results confirm the expectations of SU(3) lattice QCD of an
almost indistinguishable confining $Y-$ or $\Delta-$type at short distances
for heavy baryons. The static three-quark potential with parameters from lattice QCD does
not reproduce $bbb$ and $ccc$ spectra.

At the light of the present results a further effort to obtain a constituent quark-model
potential description of the nonperturbative lattice QCD results, as has been done in the
past for the heavy-meson systems~\cite{Qui79,Eic80,Eic08}, may help in understanding the connection between 
static three-quark potential parameters and simple Cornell-like potential descriptions. This
work is in progress~\cite{Vij15}. 

\acknowledgments
We thank to Dr. S. Meinel for valuable discussion and 
information about the present status of nonperturbative lattice QCD calculations of excited heavy 
baryon states and also for a careful reading and useful suggestions on
the manuscript. We also thank to Dr. N. Mathur for providing us with 
the numerical data of the triply charm baryon spectrum.
This work has been partially funded by
Ministerio de Educaci\'on y Ciencia and EU FEDER under 
Contracts No. FPA2010-21750 and FPA2013-47443 and by the
Spanish Consolider-Ingenio 2010 Program CPAN (CSD2007-00042) 
and by Generalitat Valenciana Prometeo/2009/129. A.V. thanks finantial 
support from the Programa Propio I of the University of Salamanca.

\end{document}